# Phase-locked array of quantum cascade lasers with an intracavity spatial filter

Lei Wang[1], Jinchuan Zhang[1]*, Zhiwei Jia[1], Yue Zhao[1], Chuanwei Liu[1], Yinghui Liu[1], Shenqiang Zhai[1], Zhuo Ning[1], Fengqi Liu[1, 2]

[1]Key Laboratory of Semiconductor Materials Science, Institute of Semiconductors, Chinese Academy of Sciences, P.O. Box 912, Beijing 100083, People's Republic of China.
[2]College of Materials Science and Opto-Electronic Technology, University of Chinese Academy of Sciences, Beijing, People's Republic of China.
*zhangjinchuan@semi.ac.cn

## Abstract

Phase-locking an array of quantum cascade lasers is an effective way to achieve higher output power and well beam quality. In this article, based on Talbot effect, we show a new-type phase-locked array of mid-infrared quantum cascade lasers with an intracavity spatial filter. The arrays show stable in-phase operation from the threshold current to full power current. We use the multi-slit Fraunhofer diffraction mode to interpret the far-field radiation profile and give a solution to get better beam quality. The maximum power is about 5 times that of a single-ridge laser for eleven-laser array device and 3 times for seven-laser array device. Considering the great modal selection ability, simple fabricating process and the potential for achieving better beam quality, this new-type phase-locked array may be a hopeful and elegant solution to get high power or beam shaping.

## Introduction

Quantum cascade lasers (QCLs) are the leading semiconductor laser sources in the mid-infrared (mid-IR) wavelength range, thanks to the features of compact size, room temperature operation, and high reliability[1, 2]. In the mid-IR region, a wide variety of applications, such as trace-gas sensing[3], breathe analysis[4], and infrared spectroscopy[5], demand the sources with high output power and well beam quality. However, the power of QCLs is limited by the narrow ridge structure (small active region size) because too wide ridge will result in poor beam quality due to generation of high-order transverse modes [6]. A hopeful solution is integrating several narrow strip lasers in parallel in an array and phase-locking all these lasers. This is called phase-locked arrays of QCLs technology. This technology is an effective solution to get higher output power and maintain well beam quality[7]. To achieve phase-locking, strong coupling among lasers in the array is needed so that all the lasers can oscillate in sync. The current coupling schemes to phase-lock QCLs include evanescent-wave coupling where adjacent lasers in the array are in a separation about a wavelength and then are coupled through the evanescent fields[8, 9], leaky wave coupling where the adjacent lasers are coupled through lateral-propagating waves coupling[10, 11], Y-coupling where several laser ridges are connected to a single waveguide to form a whole resonator[12], and antenna mutual coupling where lasers with subwavelength cavity are regarded as slot antennas and coupled with each other by mutual admittance[7]. Generally, these coupling schemes can phase-lock the array in in-phase mode (a phase shift of 0 ° between adjacent lasers) or out-of-phase mode (a phase shift of 180 ° between adjacent lasers). In-phase-mode is usually desired because of better output beam quality. However, evanescent-wave arrays often favor the out-of-phase mode because of better spatial match between pumped regions and modal distribution[8]; leaky-wave coupling arrays can operate in in-phase mode, but the modal discrimination needs very complex regrowth process to form antiguides structure[10] or need additional phase sectors that will result in serious cavity loss[11]; Y-coupling arrays have poor modal selection ability for devices with more than two branches[12]; global antenna mutual coupling scheme is only suitable for surface-emitting laser with a deep subwavelength confined cavity[7].

It is really attractive to find a simple and widely applicable way to phase-lock the array of QCLs in stable in-phase mode. Talbot effect is an optical phenomenon that the intensity pattern of an array of coherent emitters reproduces itself after a specific distance of propagation[13]. This effect has been exploited to phase-lock lasers in the near-infrared, which is called diffraction coupling scheme phase-locked array[14, 15]. Particularly, the array device with an intracavity spatial filter has demonstrated an effective solution to get the in-phase mode operation because of great modal discrimination. The modal selection principle is based on that in-phase mode and out-of-phase mode have great difference in relative location between the Talbot image and source, at the propagation of Talbot distance ($Z_t$), which is defined by[13]

$$Z_t = \frac{nd^2}{\lambda}. \qquad (1)$$

Here, n is the index of refraction; d is the center-to-center spacing of adjacent emitters; λ is the wavelength in vacuum. In this article, we realize a new-type phase-locked array of QCLs with an integrated intracavity spatial filter. Such device shows great modal selection ability and may be a hopeful solution to get high power and maintain well beam quality.

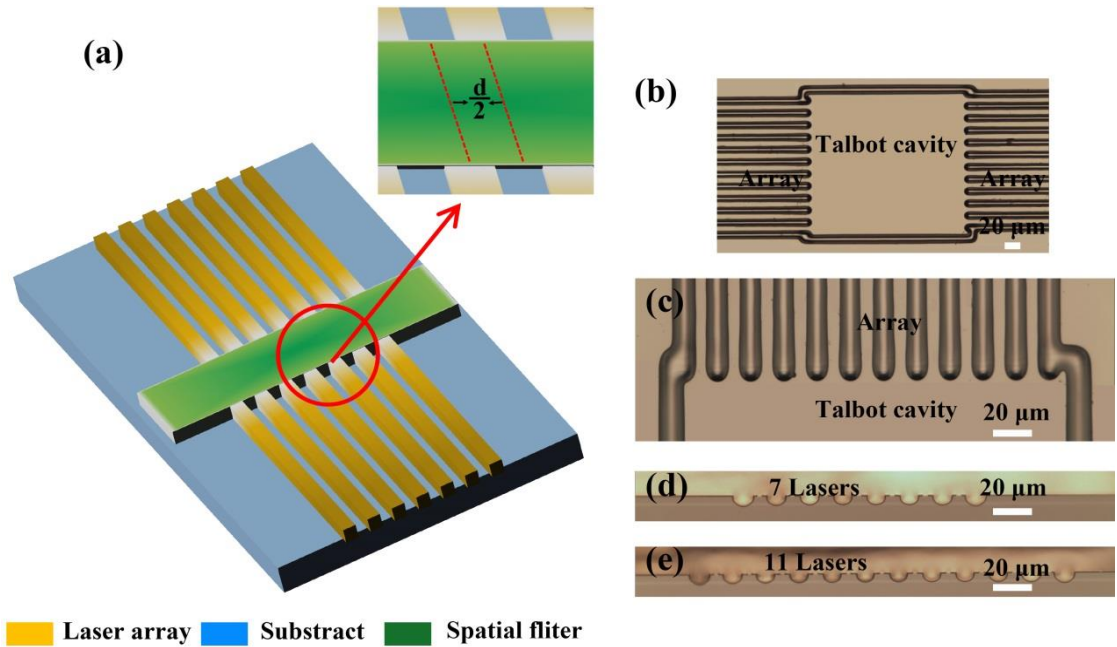

**Figure 1. Phase-locked array of QCLs with an intracavity spatial filter design.** (a) Schematic of the phase-locked array of QCLs with an intracavity spatial filter. We have fabricated devices containing seven lasers or eleven lasers in each array. (b) Optical microscope picture of the wafer after standard photolithography and wet chemical etching process. (c) Part of picture in (b) in a greater magnification. (d) Optical microscope picture of cleaved facet of a 7 laser array. (e) Optical microscope picture of cleaved facet of an 11 laser array.

## Device Design

The Fig. 1(a) shows the schematic of the device. The device contains two noncollinear laser arrays emitting at 4.6 μm, which are separated by a laterally unguided broad region (the spatial filter). The two arrays contain the same number of lasers, seven lasers or eleven lasers. The width of each laser ridge in the array is about 12 μm, so that each laser can operate in fundamental transverse mode. The center-to-center spacing of adjacent ridges in the array is 19 μm (the separation between adjacent lasers is much larger than the wavelength. This is an advantage over evanescent-wave coupling array and leaky-wave coupling array, where the separation must be within a wavelength and this is not good for heat dissipation). The length of each laser ridge is 1 mm. The structure of this device is fabricated by standard photolithography and wet chemical etching process. Fig. 1(b) and Fig. 1(c) show the optical microscope pictures of the wafer after this process. Fig. 1(d) and Fig. 1(e) show the pictures of cleaved facets of fabricated devices. They clearly show that the active region layer has been etched and the laser ridges are covered by a layer of $SiO_2$. Therefore, optical fields will be well limited in the active area and then cannot be coupled through evanescent fields or lateral-propagating waves between adjacent lasers. The phase-locking behavior and modal selection for in-phase mode can be achieved through the two special designs: (i) the length of spatial filter is equal to $Z_t$ (about 250 μm. Here, the refraction index to calculate the $Z_t$ should be the effective index in the spatial filter); (ii) as is shown in the inset of Fig. 1(a), one array is laterally offset with respect to the other by d/2.

The threshold condition of the array device can be described by the equation

$$r_1 \cdot \eta \cdot r_2 \cdot \eta \cdot \exp\left[\left(g - \alpha_w\right)\left(4L + 2L_T\right)\right] = 1, \qquad (2)$$

or

$$g = \alpha_w + \frac{1}{\left(4L + 2L_T\right)} \ln \frac{1}{r_1 \cdot r_2} + \frac{1}{\left(4L + 2L_T\right)} \ln \frac{1}{\eta^2}. \qquad (3)$$

where L is the length of ridge in the array; $L_T$ is the length of spatial filter; $r_1$ and $r_2$ are cavity facet reflectivities; $\eta$ is the coupling efficiency when the light couples into the cavity of the laser array from the spatial filter; g is the gain; $\alpha_w$ is the waveguide loss. Here, we assume that g and $\alpha_w$ in the laser array are the same as in the spatial filter. The second term in equation (3) is the mirror loss $\alpha_m$; the third term in the equation (3) is the coupling loss $\alpha_\eta$, when the light couples into the cavity of laser array from the spatial filter. Obviously, for in-phase mode and out-of-mode, there is no difference in $\alpha_w$ and $\alpha_m$. Therefore, the difference in threshold gains of different mode mainly depends on the coupling loss $\alpha_\eta$, which depends on the coupling efficiency $\eta$. The mode that has greater $\eta$ will has smaller threshold gain. As a result, the device will support this mode operation.

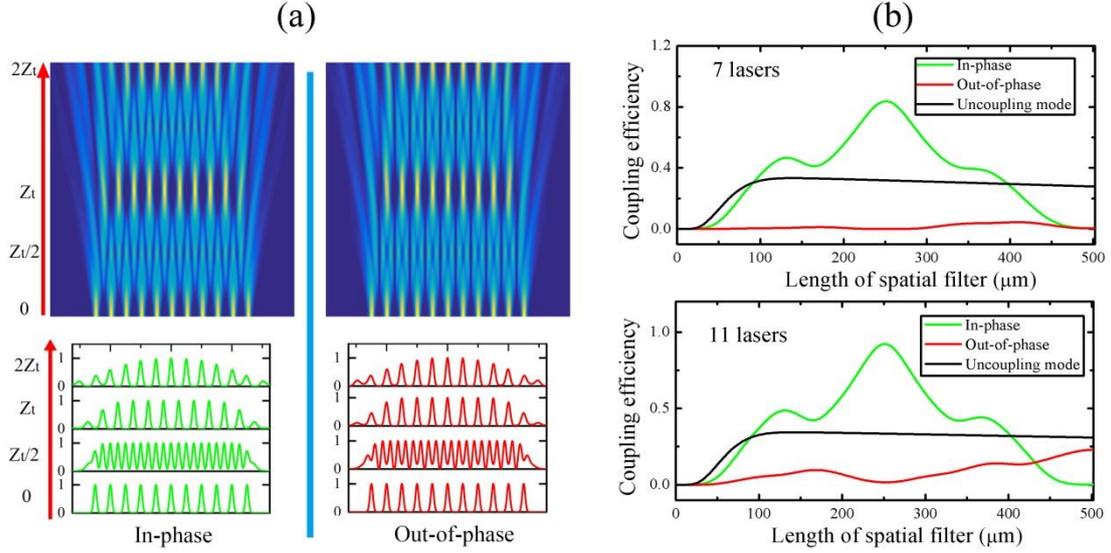

**Figure 2. Modal selection principle.** (a) Simulated intensity distributions for propagating of light emitted from the laser array in the unguided area. Here, we show the simulation for the array of 11 lasers. The upper part is the 2D intensity distributions. The propagation distance is from 0 to $2Z_t$. The under part is 1D relative intensity distributions at propagation distances of 0, $Z_t/4$, $Z_t/2$, $Z_t$, $2Z_t$. (b) Calculated coupling efficiency $\eta$ as a function of the length of spatial filter for the devices with 7 lasers in each array (upper one) and 11 lasers (under one).

The coupling efficiency $\eta$ depends on the overlap integral for the array mode profile and optical field distributions in the spatial filter near the interface between the spatial filter and the laser array. This can be described by[16]

$$\eta = \frac{\left| \int E(x,y)E_0^*(x,y)dxdy \right|^2}{\int E(x,y)E^*(x,y)dxdy \int E_0(x,y)E_0^*(x,y)dxdy}, \qquad (4)$$

where the integral plane is the interface between the spatial filter and the laser array; $E(x,y)$ is the optical field distributions in the spatial filter near the interface; $E_0(x,y)$

is the array modal distribution. To use the equation (3) to calculate the $\eta$, we use a 2D mode to simulate propagating of the light in the laterally unguided area, which is shown in Fig. 2a. At the propagation of $Z_t$, for both in-phase mode and out-of-phase mode, Talbot image (intensity distribution in unguided area) of the array mode profile (intensity distribution at the propagation of 0) is well reproduced. For out-of-phase mode, the Talbot image is aligned with respect to the array mode profile; for in-phase mode, the Talbot image is laterally offset with respect to the array mode profile by d/2. Therefore, if the two arrays are $Z_t$ away from each other and staggered by a distance of d/2, the device will suppress the out-of-phase mode, while the in-phase mode will have great coupling efficiency $\eta$ and then will be supported. Fig. 2b shows the $\eta$ as a function of the length of spatial filter $L_T$ for this device structure. When the length of spatial filter is $Z_t$ (about 250 $\mu$m), the in-phase mode will have the greatest $\eta$, which is much greater than those of out-phase mode and uncoupling mode (incoherent array). This indicates that the devices with this design will have great modal selection ability and will support the in-phase mode operation.

## Results

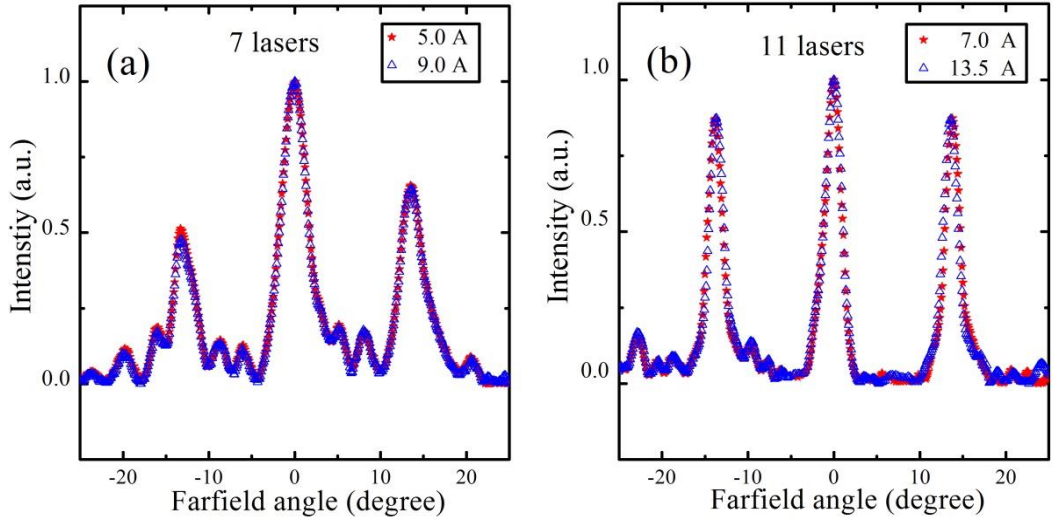

**Figure 3. Measured far-field radiation patterns.** We measure the far-field radiation patterns under different injection currents, from the threshold current to the full power current. (a) Far-field radiation pattern for seven-laser array device. (b) Far-field radiation pattern for eleven-laser array device.

**Far-field radiation pattern.** Measured far-field radiation patterns of the devices shown in Fig. 3 are of our great interest, because they reflect whether the phased-locking behavior happens and which mode operation is selected. For both seven-laser array device and eleven-laser array, the far-field radiation pattern contains three lobes, a central maximum lobe and two side lobes. Such a shape is a characteristic of in-phase mode operation[12]. The full width at half maximum (FWHM) of the central lobe is about 4 ° for seven-laser array device and 2 ° for eleven-laser array device. The interval δθ between the central maximum and one side lobe is same for seven-laser array device and eleven-laser array device (about 14 °).

We can use the multi-slit Fraunhofer diffraction mode to interpret such a far-field radiation pattern, which is given by[17]

$$F(\theta) = I(\theta) \times G(\theta), \tag{5}$$

where $F(\theta)$ is the far-field radiation pattern of the laser array; $I(\theta)$ is the far-field radiation of an individual laser in the array; $G(\theta)$ represents the multi-slit interference effect. Because the lasers in the array operate in fundamental transverse mode, the $I(\theta)$ show the shape of single-slit phenomenon. It will contain a central maximum lobe, and the FWHM of the lobe $\Delta\theta$ is about 34 °, estimated by the diffraction limit formula

$$\sin\Delta\theta = 1.22\frac{\lambda}{w}, \tag{5}$$

where w is the width of the laser ridge. The interval $\delta\theta$ is determined by $G(\theta)$, interference effect, and can be calculated by

$$d \times \sin\delta\theta = \lambda. \tag{6}$$

The calculated result of $\delta\theta$ is about 14 °, which is the same as the experimental result. Because $\Delta\theta$ is about 34 °and $\delta\theta$ is about 14 °, three interference peaks are in the range of the central lobe of $I(\theta)$. That is why the measured far-field radiation contains three lobes.

The measured far-field radiation patterns show that the devices can provide coherent beam with smaller divergence than that of a single-ridge laser. Furthermore, we can achieve better beam equality (single-lobe far-field profile), decreasing d to increase the $\delta\theta$ or increasing w to decrease the FWHM of $I(\theta)$ to ensure that only one interference peak can be contained in the range of FWHM of I(θ). The FWHM of the central lobe in measure far-field radiation patterns is very small. This indicates that such device may be a potential method for beam shaping, if a single-lobe far-field radiation is achieved.

Besides, for both devices, the in-phase operation shows great modal stability under different injection currents, from the threshold current to the full power current.

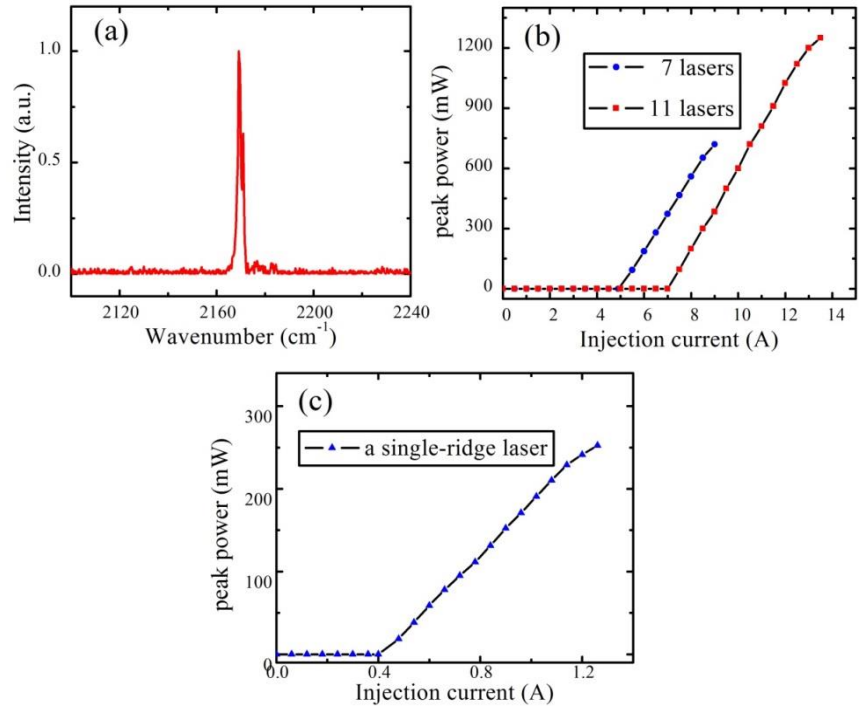

**Figure 4. Measured spectrum and power of the devices.** All the tests were under pulse operation (1% duty cycle with 600 ns pulse width). We kept the temperature of heat sink 283K. (a) Emission spectrum of the array devices. (b) P-I curves for 7 laser array device and 11 laser array device. (c) P-I curves for a single-ridge laser.

**Spectrum and power.** Fig. 4a shows the measured emission spectrum. The peak wavelength is 4.6 μm (about 2174 cm$^{-1}$). Fig. 4b shows optical power-current (P-I) characterization of uncoated devices in pulse operation (1% duty cycle with 600 ns pulse width). The temperature of heat sink was kept 283 K. The eleven-laser array device exhibits a maximum power of 1250 mW, a slope efficiency of 0.19 W/A, and a threshold current density of 1.9 kA/cm$^2$ (threshold current is 7.0 A; current injection area is about $3.7 \times 10^{-3}$ cm$^2$); the 7 laser array device exhibits a maximum power of 780 mW, a slope efficiency of 0.15 W/A, and a threshold current density of 2.0 kA/cm$^2$ (threshold current is 4.9 A; current injection area is about $2.4 \times 10^{-3}$ cm$^2$). The eleven-laser array device shows lightly smaller threshold current density and greater slope efficiency. The possible reason may be that the edge diffraction loss has smaller influence for array with more emitters. We also measured P-I curve of a normal single-ridge laser (the laser ridge is 12-μm-wide and 2-mm-long) for comparison. The single-ridge device shows a maximum power of 250 mW, a slope efficiency of 0.29 W/A and a threshold current density of 1.6 kA/cm$^2$ (threshold current is 400 mA; current injection area is about $2.5 \times 10^{-4}$ cm$^2$). It can be found that the integrated spatial filter could result in slightly increase in threshold current density and obviously decrease in slope efficiency. As a result, the maximum power is just about 5 times that of a single-ridge laser for eleven-laser array device and 3 times for seven-laser array device.

Device performance degeneration for the individual laser in the array is a result of additional optical loss due to the existence of spatial filter. Although the calculated coupling efficiency η in Fig 2b is nearly 90%, fabricating the spatial filer will bring other optical loss that will result in device performance degeneration. From the Fig. 1c, we see that the laser ridges near the spatial filter show taper shape. Such shape is formed because of nonuniform distribution of etching solutions in this area during the wet chemical etching process. Such Taper shape will lead to serious waveguide losses and therefore the devices performance will degrade. To avoid this problem, we can use dry etching technique to fabricate the devices with straight laser ridge.

## Discussion

In conclusion, we realize a new-type phase-locked array of quantum cascade lasers, by integrating a spatial filter (laterally unguided area) between two noncollinear laser arrays. The device design is based on Talbot effect. At propagation of $Z_t$, the Talbot images of in-phase mode and out-phase-mode have distinct relative location with respect to the source. Therefore, to ensure great modal discrimination to support the in-phase mode operation, the length of spatial filter is $Z_t$ and two laser arrays are laterally offset by d/2. The measure far-field radiation patterns reflect stable in-phase mode operation under different injection currents, from threshold current to full power current. The far-field radiation pattern contains three lobes, a central maximum lobe and two side lobes. The interval between central lobe and side lobe is about 14 °. This indicates that the beam divergence of such array device is smaller than that of a single-ridge laser. We can use the multi-slit Fraunhofer mode to interpret the shape of far-field radiation pattern. Better beam quality, a single-lobe far-field profile, can be obtained through decreasing the center-to-center spacing of adjacent lasers or increasing the width of laser ridge. The 11-laser array device shows a maximum power of 5 times that of a single-ridge laser; the 7-laser array device shows a maximum power of 3 times that of a single-ridge laser. The device performance degeneration of individual laser in the array may be caused by the taper shape of laser ridge near the spatial filter. This may be avoided by using dry etching technique to fabricate the devices with straight laser ridge. Considering the great modal selection ability, simple fabricating process and the potential for achieving better beam quality, this new-type phase-locked array may be a hopeful and elegant solution to get high power or well beam quality.

## Methods

**Optical simulation and calculation.** Every laser in the array operates in fundamental transverse mode. Radiation from each emitter is characterized by Gaussian beam with the waist equally to the width of laser ridge. There is no residual reflection from the interface between the laser ridges. Coupling efficiency of each array mode is calculated according to equation (3). As for uncoupled mode (incoherent arrays), we calculated coupling efficiency of each individual laser in the array and then make an average.

**Growth and fabrication.** The QCL wafer was grown an n-InP substrate (Si-doped, $2\times10^{17}$ cm$^{-3}$) by solid-source molecular beam epitaxy (MBE). The materials structure is based on a two-phonon resonance design for emission for 4.6 μm. The active core includes 30 stages of strain-compensated In$_{0.669}$Ga$_{0.331}$As/In$_{0.362}$Al$_{0.638}$As quantum wells and barriers. The entire structure was as follows: 1.2 μm lower cladding (Si-doped InP, $2.2\times10^{16}$ cm$^{-3}$), 0.3 μm n-In$_{0.53}$Ga$_{0.47}$As layer (Si-doped, $4\times10^{16}$ cm$^{-3}$), 30 active/injector stages, 0.3 μm n-In$_{0.53}$Ga$_{0.47}$As layer (Si-doped, $4\times10^{16}$ cm$^{-3}$), 2.4 μm upper cladding (Si-doped InP, $2.2\times10^{16}$ cm$^{-3}$) and 0.6 μm cap cladding (Si, $1\times10^{19}$ cm$^{-3}$). The layer sequence of one period, starting from the injection barrier is as follows (thickness in nanometers):

**3.8**/1.2/**1.3**/4.3/**1.3**/3.8/**1.4**/3.6/**2.2**/2.8/**1.7**/2.5/**1.8**/2.2/<u>**1.9**</u>/<u>2.1</u>/<u>**2.1**</u>/<u>2.0</u>/**2.1**/1.8/**2.7**/1.8, In$_{0.362}$Al$_{0.638}$As barrier layers is in bold, In$_{0.669}$Ga$_{0.331}$As quantum well layers are in normal font, and doped layers (Si-doped, $1.5\times10^{17}$ cm$^{-3}$) are underlined. The waveguide structure was fabricated by standard photolithography and wet chemical etching process. Then a 500 nm-thick-SiO$_2$ was deposited on the top by Plasma Enhanced Chemical Vapor Deposition (PECVD) as electrical insulating layer. 2-μm-wide windows were opened on the top of laser ridges and Talbot cavity. Then a Ti/Au layer was evaporated on the top as electrical contact, and a 3-μm-thick Au layer was electroplated on the electrical contact to further improve the heat dissipation. After the sample was thinned down to 100 μm, the back electrical contact was provided by a deposited Ge/Au/Ni Au metal contact layer. Finally, the wafer was cleaved into separate devices.

**Device testing.** The devices were mounted epilayer side down on copper heat sinks with In solder for testing. The temperature of heat sinks was kept at 283 K by a thermoelectrically cooled holder. Pulsed operation was taken at 1% duty cycle with 600 ns pulse width. The output optical power was measured with a calibrated thermopile detector placed in the front of the cavity surface of the lasers. The emission spectrum was measured by a Fourier transform infrared (FTIR) spectrometer in rapid scan mode with a resolution of 0.2 $cm^{-1}$. Devices was mounted on a computer controlled rotational stage with a step resolution of 0.05 ° for far-field testing, and the light was detected by a nitrogen-cooled HgCdTe detector placed 30 cm away from the cavity surface of lasers.

## Acknowledgements

This work was supported by the National Basic Research Program of China (Grant Nos. 2013CB632801), National Key Research and Development Program (Grant Nos. 2016YFB0402303), National Natural Science Foundation of China (Grant Nos. 61435014, 61627822, 61574136, 61306058), Key Projects of Chinese Academy of Sciences (Grant No. ZDRW-XH-2016-4) and Beijing Natural Science Foundation (Grant No.4162060).

## Author Contributions

L.W. came up with the idea, designed the waveguide structure, fabricated the devices, performed the testing, and wrote the manuscript. J.Z. told about the idea with L.W.

and revised the manuscript. Z.J. conducted growth of InP waveguides. Y.Z. took part in devices testing. C.L. and H.L. took part in devices fabrication. S.Z conducted the active region materials growth. N.Z. design the energy structure of QCLs. F.L. supervised the project.

## Additional Information

**Competing financial interests:** The authors declare no competing financial interests.